# Surface terminations and layer-resolved spectroscopy in 122 iron pnictide superconductors

Ang Li,[1,3,9*] Jiaxin Yin,[2,1] Jihui Wang,[1] Zheng Wu,[1] Jihua Ma,[1,4] Athena S. Sefat,[5] Brian C. Sales,[5] David G. Mandrus,[5] Rongying Jin,[6] Chenglin Zhang,[7] Pengcheng Dai,[7] Bing Lv,[1] Xuejin Liang,[2] P.-H. Hor,[1] C.-S. Ting[1] and Shuheng H. Pan[2,1,8*]

[1]Department of Physics & Texas Center for Superconductivity, University of Houston, Houston, Texas 77004, USA

[2]Institute of Physics, Chinese Academy of Sciences, Beijing 100190, China

[3]State Key Laboratory of Functional Materials for Informatics, Shanghai Institute of Microsystem and Information Technology, Chinese Academy of Sciences, Shanghai 200050, China

[4]Department of Physics, Boston College, Chestnut Hill, Massachusetts 02467, USA

[5]Materials Science & Technology Division, Oak Ridge National Laboratory, Oak Ridge, Tennessee 37831, USA

[6]Department of Physics & Astronomy, Louisiana State University, Baton Rouge, Louisiana 70803, USA

[7]Department of Physics and Astronomy, Rice University, Houston, Texas 77005, USA

[8]Collaborative Innovation Center of Quantum Matter, Beijing, China

[9]CAS-Shanghai Science Research Center, 239 Zhangheng Road, Shanghai 201203, China

*Corresponding authors: angli@mail.sim.ac.cn; span@iphy.ac.cn

Abstract

The surface terminations of 122-type alkaline earth metal iron pnictides $AE$Fe$_2$As$_2$ ($AE$ = Ca, Ba) are investigated with scanning tunneling microscopy/spectroscopy (STM/STS). Cleaving these crystals at a cryogenic temperature yields a large majority of terminations with atomically resolved ($\sqrt{2}\times\sqrt{2}$)R45 or 1×2 lattice, as well as the very rare terminations with 1×1 symmetry. By means of lattice alignment and chemical marking, we identify these terminations as ($\sqrt{2}\times\sqrt{2}$)R45-$AE$, 1×2-As, and ($\sqrt{2}\times\sqrt{2}$)R45-Fe surfaces, respectively. Layer-resolved spectroscopy on these terminating surfaces reveals a well-defined superconducting gap on the As terminations, while the gap features become weaker and absent on $AE$ and Fe terminations respectively. The local gap features are hardly affected by the surface reconstruction on As or $AE$ surface, whereas a suppression of them along with the in-gap states can be induced by As vacancies. The emergence of two impurity resonance peaks at ±2 meV is consistent with the sign-reversal pairing symmetry. The definite identification of surface terminations and their spectroscopic signatures shall provide a more comprehensive understanding of the high-temperature superconductivity in multilayered iron pnictides.

## I. INTRODUCTION

The discovery of iron-based superconductors marked a significant progress in the study of high-temperature superconductivity [1-3]. The iron pnictides (chalcogenides) are characterized by a multilayered crystal structure with Fe-Fe planes as the common ingredient. It is generally believed that the superconductivity develops in the Fe plane and all five Fe *d* orbitals contribute to the low-energy physics, leading to a multi-band nature of this class of materials [4]. A direct characterization of the electronic properties of each building layer is thus essential for unraveling the key element(s) governing the superconductivity in such a complex system. STM/STS are perfect tools for extracting structural and spectroscopic information down to atomic level. They have played a critical role in exploring the exotic orders in iron-based superconductors [5]. So far, iron-based superconductors like Fe(Se,Te) and LiFeAs have been relatively well characterized by STM/STS due to their definite cleavage, while the study of *AE*122 compounds suffers from the controversial identification of the cleavage planes and the complex surface reconstructions [6-23]. Therefore, clarification of the cleavage, identification of the resulting terminations, and subsequent atomic-layer-resolved spectroscopic investigations become crucial and highly demanded for studying the *AE*122 iron-based superconductors.

Among iron-based superconductors, the *AE*122 compounds are superior for their large crystal size and widely accessible chemical doping range [2,3]. They have a layered crystal structure with the weakest bonding between the adjacent *AE* and As layers (Fig. 1(a)). Accordingly, these crystals tend to cleave there between and the *AE* and As planes are most likely to be exposed as the surface terminations. Due to the polarized charge distribution among different layers in this type of materials, structural and/or electronic reconstructions are spontaneous at these terminating surfaces. Consequently, the modified surface morphologies make the count of atoms and their identification more complicated. Based on the early STM studies [6-23], two major assignments have been proposed for the *AE*122-crystal surface terminations: (1) Upon cleaving, half of *AE* atoms redistribute uniformly on each surface termination to form a ($\sqrt{2}\times\sqrt{2}$)R45 or 1×2 superstructure respectively [6-9,12,14,19-23]; (2) the *AE*-layer collapses and two complete As layers are exposed instead in ($\sqrt{2}\times\sqrt{2}$)R45 or 1×2 format [10,11,13,18]. In this article we present high-resolution STM/STS results on a series of cryogenically cleaved $AE$Fe$_2$As$_2$ (Ba(Fe$_{1-x}$Co$_x$)$_2$As$_2$, Ba$_{1-x}$K$_x$Fe$_2$As$_2$, and CaFe$_2$As$_2$) single crystals. We demonstrate that three types of ordered surface structure can be distinguished as (1) ($\sqrt{2}\times\sqrt{2}$)R45 reconstruction of the complete *AE* lattice, (2) 1×2 stripes from the As dimerization, and (3) the rarely encountered ($\sqrt{2}\times\sqrt{2}$)R45 pattern of Fe lattice. Differential conductance measurements show superconducting gap features in the low-energy excitation spectra on both *AE* and As terminations, while not in that on the Fe exposure. Such spectroscopic discrepancies indicate that the As-Fe-As tri-layer block is essential to the superconductivity in iron pnictides hence the key unit to explore.

## II. EXPERIMENTAL METHOD

The single crystalline samples of $BaFe_2As_2$, $SrFe_2As_2$ and $CaFe_2As_2$ with various doping were grown using the self-flux method [24-26]. The STM/STS experiments were carried out on a home-built ultra-high vacuum low-temperature STM. Samples were cleaved *in situ* below 30 K and immediately transferred to the STM head which was already at the base temperature of 4.3 K. The scan tips were prepared from polycrystalline tungsten wires by electrochemical etching and subsequent field-emission cleaning. Topographic images were acquired in the constant-current mode with the bias voltage applied to the sample. Differential conductance spectra were recorded with the standard lock-in technique. When describing the lattice symmetry, we choose the $ThCr_2Si_2$-type tetragonal notation throughout this article so that the low-temperature orthorhombic symmetry [27] can be denoted as "($\sqrt{2}\times\sqrt{2}$)R45", or "rt2" for short.

## III. RESULTS AND DISCUSSION

The most commonly observed surface terminations in $AE$Fe$_2$As$_2$ are shown in Fig. 1(b) and Fig. 1(c). The 1×2 superstructure in Fig. 1(b) consists of one-dimensional stripes with inter-stripe distance ∼8 Å, twice the tetragonal lattice constant. Along the stripe are grains at 4 Å spacing and within each grain two atoms can be resolved to form a dimer (inset in Fig. 1(b)). This dimerization can switch its direction by 90° thus "twins" (not crystallographic twins) are quite often found on a striped surface as shown in Fig. 1(b). Apparently the dimerization is a surface phenomenon as no bulk evidence has been reported from diffraction experiments. Its structural nature is further evidenced by the independence of topographic image on the bias voltage. The second type termination in Fig. 1(c) exhibits a bias-independent square-like lattice with much smaller corrugation. The unit cell is enlarged by $\sqrt{2}\times\sqrt{2}$ times from the tetragonal and orients at 45° to the 1×2 stripes (Fig. 2(a)). At first sight only half of the atoms are resolved if one attempts to assign them to the *AE* or As layer. More careful examination, however, reveals the complete atomic coverage in a buckled way (inset in Fig.1(c)). Such a rt2 lattice is widely observed in Ba(Fe$_{1-x}$Co$_x$)$_2$As$_2$ including the heavily overdoped side, where the bulk low-temperature orthorhombic phase is completely suppressed [28]. Accordingly the rt2 superstructure represents another type of surface reconstruction. In addition to 1×2 and rt2, a third type surface morphology with distinctive 1×1 lattice symmetry (Fig. 1(d)) is very occasionally observed. The statistics strongly suggests an unusual origin of the 1×1 structure. In other words, the common 1×2 and rt2 morphologies are expected to be the natural form of *AE* and/or As exposed after cleaving.

It is hard to decide the chemical identity of the 1×2 or rt2 reconstructed surface as the *AE* and As layers share the same lattice symmetry in the bulk. Moreover, the apparent height difference between rt2 and 1×2 regions is noticeably small (~0.07 nm for $BaFe_2As_2$ and ~0.01 nm for $CaFe_2As_2$) as compared to the crystalline *AE*-As interlayer distance (0.19 nm [27]). Due to the fact that the constant current topographic image convolutes the spatial variation of the integrated local density of states (LDOS) and the geometrical corrugations, we cannot simply assign these terminations by their apparent height. Nevertheless, the rt2 and 1×2 surfaces should either belong to the same *AE*/As termination or correspond to the upper and lower terrace of a monoatomic *AE*-As step [29]. A critical clue comes from the in-plane lattice alignment between 1×2 and rt2. In the former case, the centerline of each stripe should point to the rt2 superlattice with an offset of half a unit cell, as schematically drawn in Fig. 2(b). For the *AE*-As monoatomic step, on the contrary, 1×2 stripes would line up with the rt2 superlattice (Fig. 2(c)). The STM images in Fig. 2(a) and the insets are clearly consistent with the step picture. Hence *AE* and As each contribute to one of the commonly observed surface terminations with specific structural reconstruction. The reduced step height can be explained by *AE*'s much smaller contribution to the density of states near Fermi level [30].

To further nail down the identities of rt2 and 1×2 terminations, selectively marking one of them with known dopants would be simple and straightforward. Here we chose the potassium doped $BaFe_2As_2$ in which K dopants partially substitute Ba. Similar to the report in Ref. 20, bright and dark sites are readily visible at atomic level in the rt2 topography (Fig. 2 (d) and (e)), while such contrast is absent in the 1×2 stripes. The percentage of dark atoms is approximately 40%, agrees fairly well with the nominal K concentration. The large population and good agreement with doping content lead us to the conclusion that two major surface terminations are created by cold cleaving between *AE* and As planes; the exposed *AE* layer buckles to form rt2-superstructure and the As atoms in arsenic plane dimerize into one-dimensional stripes. They each do not necessarily cover the entire cleaving surface. Instead a rough 50%-50% mix of them is found separated by atomic steps. The surface reconstructions and their patchy distribution do not conflict with the requirement for charge balance between *AE* and As layers.

The 1×1 termination (Fig. 1(d) and Fig. 2(f)) exhibits the same lattice symmetry as that of the bulk *AE* and As. Considering its very low probability of occurrence, it could be an un-reconstructed version of the *AE* or As exposure [22,31]. In spite of that, the 1×1 symmetry may reflect the (√2×√2)R45 pattern of Fe lattice as well. The rt2 pattern of Fe, when surrounded by As stripes, will have a unique atomic registration with the As stripes from two orthogonal directions as illustrated in Fig. 2(g) and (h). Note that there is a half-unit-cell shift in the alignment of rt2 Fe with respect to the As stripes from *a* and

*b* directions. The height profiles along two lines in Fig. 2(f) show clear evidence for such a phase shift; hence prove the 1×1 surface to be rt2 pattern of Fe layer. We emphasize that this anisotropic atomic arrangement is exclusive for rt2 Fe. Although the Fe termination is very rare, it offers a perfect chance to evaluate the electronic properties of each building layer in iron pnictide superconductors.

In Fig. 3 the spatially averaged differential conductance spectra on optimally doped Ba(Fe,Co)$_2$As$_2$ ($T_C$ = 22 K) show the superconducting energy gap on both Ba and As terminations with the gap magnitude $\Delta$ = 6 meV (one half the distance between coherence peaks). The ratio $2\Delta/k_BT_C$ = 6.3, underlining a strong coupling superconducting state. Interestingly, the gap features are well defined in 1×2 As surface, while the spectrum on Ba termination exhibits reduced coherence peaks and more in-gap states near the Fermi level. The cause of such spectral differences is still an open question, and we try to understand it from the orbital perspective in our following paper [32]. In strong contrast, the Fe termination is characterized by an overall V-shaped spectrum without clear superconducting gap feature. These findings highlight the role of As layer in electronic pairing and demonstrate that superconductivity emerges from the As-Fe-As tri-layer block whose integrity is vital. Indeed, many of the system's magnetic and electronic properties are known to be very sensitive to the Fe-As height [33,34]. Removing the As layer can dramatically alter the local atomic environment of Fe hence the electronic pairing.

In addition to the chemical identity, the lattice reconstruction can also influence the surface electronic structure by inducing band folding in the momentum space [35]. The representative broken symmetry in *AE* and As terminations, however, does not alter the low-energy LDOS significantly. For example, no perceivable change in the gap size is detected with the 1×2 periodicity (Fig. 4(a) and (b)), except that the width of coherence peaks is very weakly modulated across the stripe (Fig. 4(d)). This is consistent with the fact that the superconducting coherence length (typically more than 2 nm [7,16]) is much larger than the periodicities of these superlattices.

To gain further insight into the superconducting pairing mechanism, we now study the LDOS near a single atomic impurity. We focus on the distinct dimer vacancy on an As surface, which termination has the most well-defined superconducting spectral features. Fig. 5 shows the differential conductance spectrum taken at the center of such a defect. When compared with the reference spectrum (measured tens of nm away from any defects), the coherence peaks are strongly suppressed and the low-energy LDOS inside the gap is significantly enhanced. By subtracting the reference spectrum from that of the As vacancy, as shown in the right inset of Fig. 5, we can identify a pair of in-gap bound states symmetrically located at ±2 meV. The As atoms are known to bridge the electron hopping between the next nearest neighboring irons. The missing As atoms effectively impose a scattering potential on the iron site right underneath. Interband impurity scattering experiencing a phase change in the superconducting order parameter would give rise to a pair of resonance peaks symmetrically located with respect to the Fermi energy, while no such states for the non-phase-changing superconductor [36]. Hence the observed in-gap resonances are consistent with a sign-reversal pairing symmetry.

## IV. SUMMARY

We have performed an STM/STS study on the *AE*-122 iron pnictides. Cleaving at a cryogenic temperature creates the predominant rt2-buckled *AE* and 1×2-dimerized As terminations. A complete *AE* and As atomic coverage is confirmed in each termination. Very occasionally the crystal cleaves between the As and Fe layers leaving the rt2 pattern of Fe exposed with characteristic 1×1 lattice symmetry. The superconducting energy gap is observed on *AE* and As terminations, while no such features are found on the Fe termination. Our atomic-layer-resolved spectroscopic study suggests that the As-Fe-As tri-layer block is essential for the development of superconductivity in 122 pnictides. Locally the surface reconstructions have very limited impacts on the superconducting gap on *AE* and As terminations. Local pair-breaking is observed on the As vacancy that is consistent with the potential scattering in an s±-wave superconductor. The definite identification of various terminating surfaces and detailed spectroscopic characterization provide us valuable information towards a comprehensive understanding of the iron-based superconductivity.

**Acknowledgments**

We sincerely thank Donghui Lu and Zhongyi Lu for helpful discussions. This work is supported by State of Texas through TcSUH, Chinese Academy of Sciences, NSFC (11227902, 11322432, 11190020), the Strategic Priority Research Program B (XDB04040300, XDB07000000) and the Hundred Talents Program of the Chinese Academy of Sciences, Ministry of Science and Technology of China (2012CB933000, 2012CB821400, 2015CB921300), and U.S. DOE (DE-SC0012311).

Figures

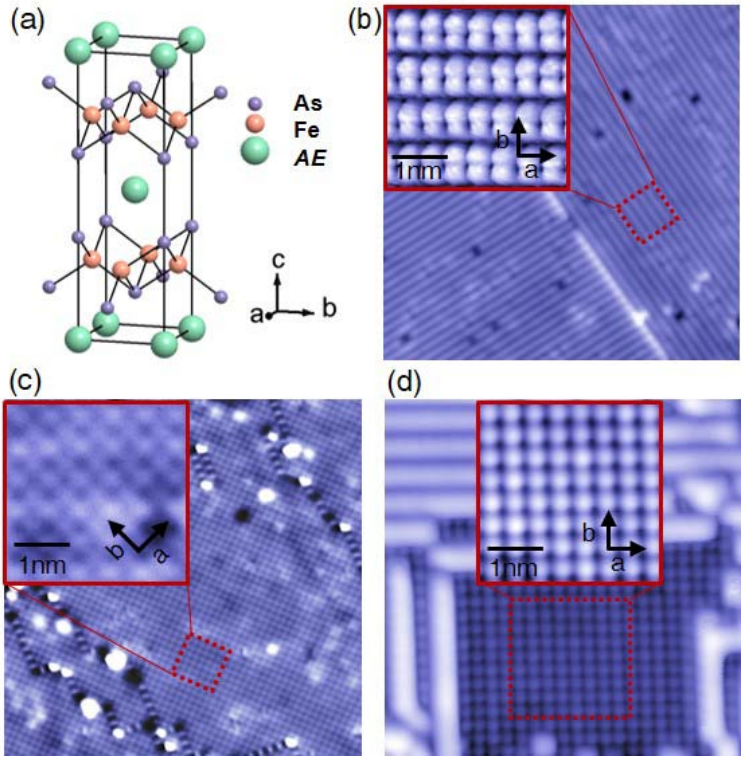

FIG. 1: (a) Schematic crystal structure of $AE$Fe$_2$As$_2$. (b-d) Surface morphologies demonstrating (b) 1×2 (V = 100 mV, I = 30 pA), (c) rt2 (V = 20 mV, I = 2 nA), and (d) 1×1 (V = 50 mV, I = 2 nA) structures. Insets in (b-d) are the zoom-in image of 1×2 dimers (V = 20 mV, I = 8 nA), rt2 buckling (V = 20 mV, I = 2 nA) and 1×1 surface (V = 50 mV, I = 2 nA), respectively.

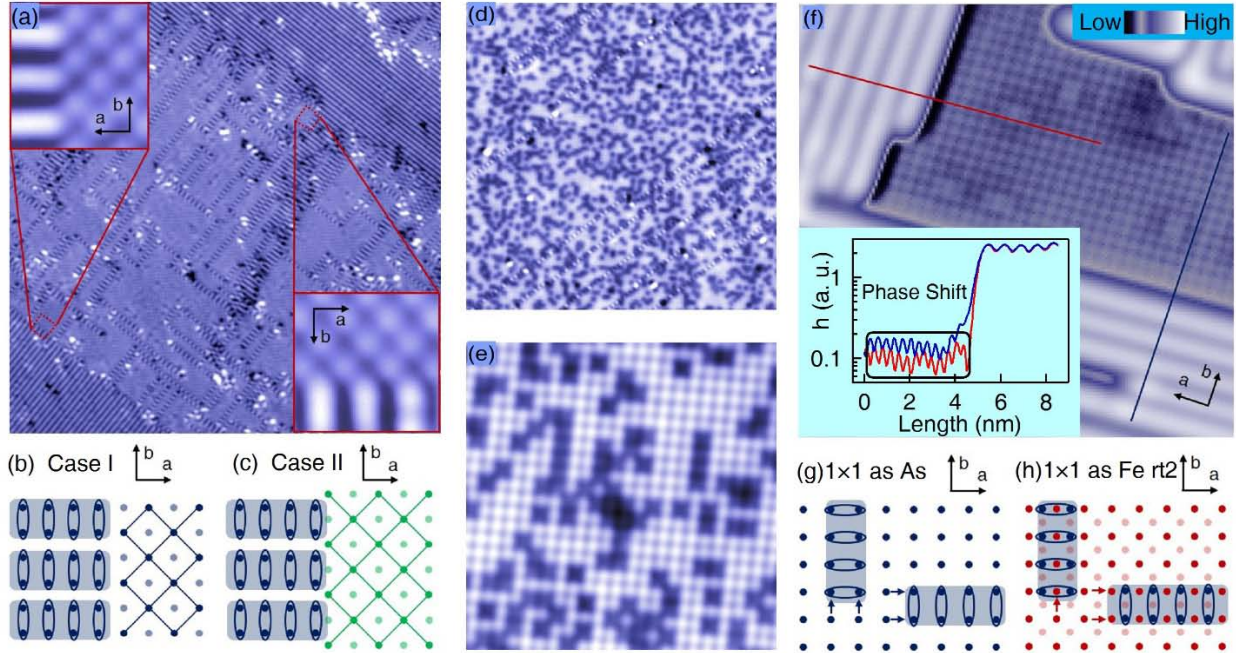

FIG. 2: (a) The joint area between rt2 and 1×2 in CaFe$_2$As$_2$ (70×70 nm$^2$, V = 50 mV, I = 1 nA). Insets are the zoom-in images of boundaries along two orthogonal directions. (b) and (c) Schematic drawings for the in-plane atomic arrangement of 1×2 and rt2 when they belong to the same (Case I, either As or *AE*) or two different atomic layers (Case II). Ellipses and solid lines represent the dimers and rt2 superlattice respectively. The rt2 topographies of Ba$_{0.6}$K$_{0.4}$Fe$_2$As$_2$ are shown in (d) 50×50 nm$^2$, V = -100 mV, I = 2 nA and (e) 12.5×12.5 nm$^2$, V = 100 mV, I = 2 nA. (f) 1×1 area surrounded by the 1×2 stripes in CaFe$_2$As$_2$ (13×13 nm$^2$, V = 50 mV, I = 1 nA). Inset: height profiles along two orthogonal lines indicated in the main panel. Once the stripe corrugations are aligned from the right, there is a half-unit-cell phase shift in the 1×1 profile on the left. (g) and (h) Schematic drawings for the in-plane atomic arrangement of stripes vs 1×1 lattice assuming: (g) they both belong to the As plane; (h) 1×1 is rt2-buckled Fe (As: dark blue, Fe: dark and light red).

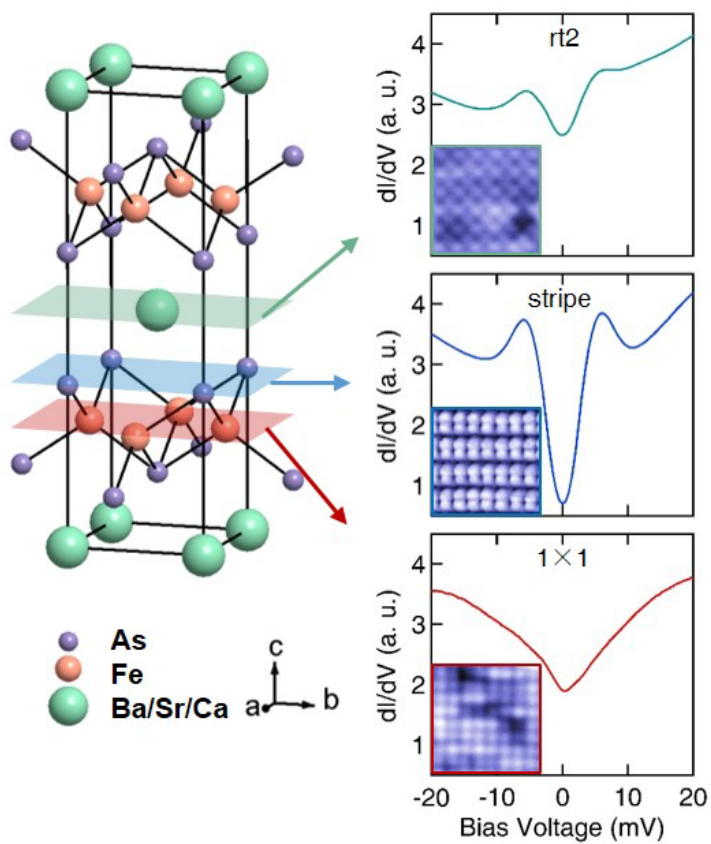

FIG. 3: Spatially averaged differential conductance spectra on three surface terminations in optimally doped Ba(Fe,Co)$_2$As$_2$ (V = -20 mV, I = 0.67 nA).

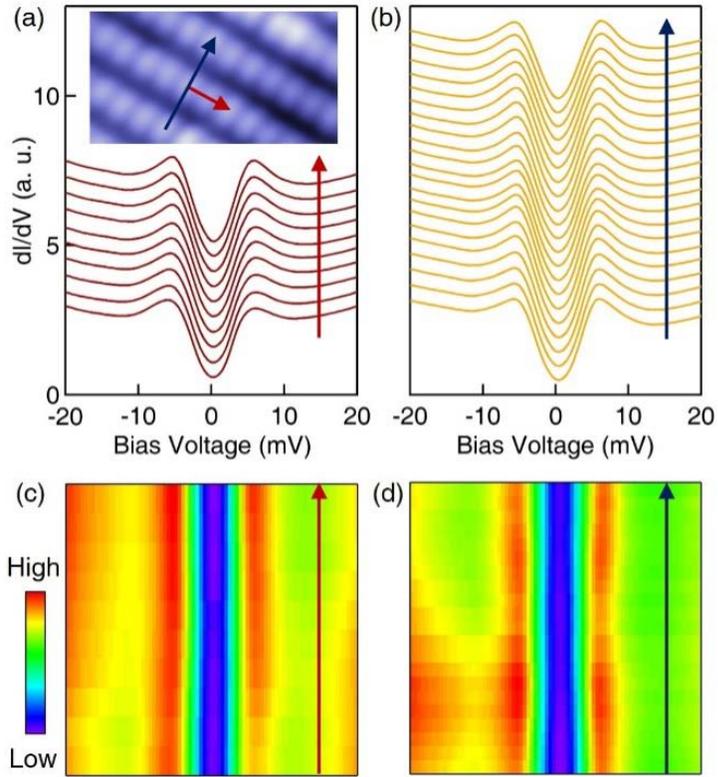

FIG. 4: (a) and (b) Series of differential conductance spectra measured along and across the As stripes for optimally doped Ba(Fe,Co)$_2$As$_2$ (V = -20 mV and I = 0.67 nA). The trajectories are drawn in the inset of (a). Spectra are offset for clarity. (c) and (d) Intensity plot of the spectra in (a) and (b) respectively.

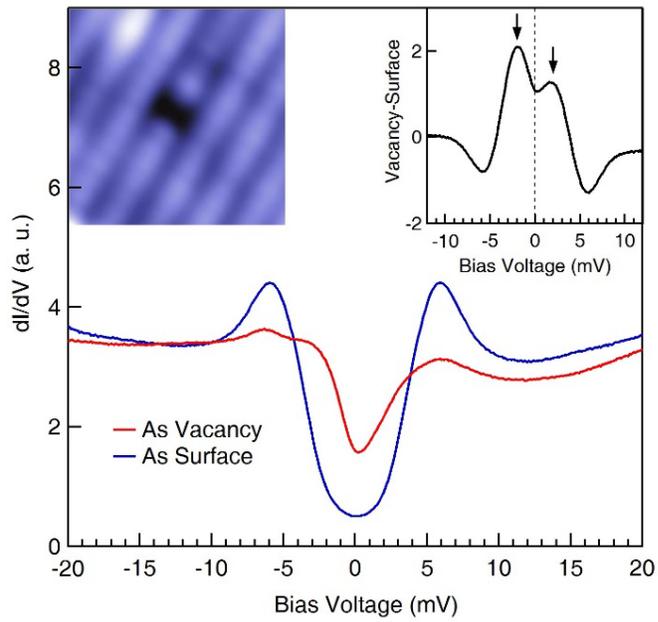

FIG. 5: Differential conductance spectra measured at the center of an As dimmer vacancy (red) and on the regular As surface (blue) in optimally doped Ba(Fe,Co)$_2$As$_2$ (V = -20 mV and I = 0.67 nA). Left inset: topographic image; right inset: the difference spectrum (see text). A pair of LDOS peaks at ± 2 mV are marked with arrows.